\documentclass{jaa}
\usepackage{natbib}
\usepackage{graphicx}	
\usepackage{amsmath}	
\usepackage{amssymb}	
\usepackage{url}
\usepackage{float}
\usepackage{mathrsfs}
\usepackage{mathtools}
\usepackage{siunitx}
\usepackage{bigints}
\usepackage{longtable,multirow}
\usepackage{tabularx}
\usepackage{footnote}
\usepackage{tablefootnote}
\usepackage{subfig}
\usepackage{placeins}
\usepackage{listings}
\usepackage{color,hyperref}
\usepackage{gensymb}
\usepackage{pdflscape}
\usepackage{verbatim}
\usepackage{textgreek}
\usepackage{newtxtext}                  
\usepackage[slantedGreek]{newtxmath}
\newcommand{\ignore}[1]{}

\newcommand{\nus}{\emph{NuSTAR}}
\newcommand{\lxp}{\emph{AstroSat}/LAXPC}
\newcommand{\lxpp}{\emph{AstroSat}-LAXPC20}

\newcommand{\astrosat}{\emph{AstroSat}}
\newcommand{\rxte}{\emph{RXTE}}

\newcommand{\lmxb}{4U 1636\ensuremath{-}536}

\begin{document}\sloppy

\title{\nus\ and \astrosat\ observations of thermonuclear X-ray bursts with short-recurrence times in \lmxb\ }


\author{Pinaki Roy\textsuperscript{1,*}, Aru Beri\textsuperscript{1,2} and Aditya S. Mondal\textsuperscript{3}}
\affilOne{\textsuperscript{1}Department of Physical Sciences, IISER Mohali, Punjab 140306, India\\}
\affilTwo{\textsuperscript{2}School of Physics and Astronomy, University of Southampton, Hampshire, SO17 1BJ United Kingdom\\}
\affilThree{\textsuperscript{3}Department of physics, Visva-Bharati, Santiniketan, West Bengal 731235, India\\}


\twocolumn[{

\maketitle

\corres{pinaki.roy1989@gmail.com}

\msinfo{}{}

\begin{abstract}
We report results from the spectro-timing analysis of the atoll source \lmxb\ observed with \nus\ and \astrosat\ during its hard spectral state. In three observations of 207 ks total exposure, we identify 31 thermonuclear X-ray bursts including five doublets and a triplet. Recurrence time as short as 3.8 minutes is seen in one of the doublets. To the best of our knowledge, this is the shortest recurrence time known for this source. Our time-averaged spectroscopy during the bursts indicate the presence of an additional powerlaw or a blackbody component in a few cases, perhaps due to varying temperatures during bursts or plausible deviation from ideal blackbody behavior, however, it is difficult to probe this using the time-resolved spectroscopy owing to limited statistics. Time-resolved bursts are well fit using an absorbed blackbody model with temperatures varying between 1.7 and 2.2 keV. Burst oscillations around 581 Hz are detected with 3$\sigma$ confidence during the decay phase in two of the X-ray bursts. One of the burst oscillations is seen at 582 Hz, a frequency observed during the 2001 superburst in this source.
\end{abstract}


}]


\doinum{}
\artcitid{\#\#\#\#}
\volnum{000}
\year{0000}
\pgrange{1--}
\setcounter{page}{1}
\lp{1}

\section{Introduction}
Thermonuclear (Type I) X-ray bursts are sudden surges in X-ray emission from an accreting neutron star in a Low-Mass X-ray Binary (LMXB). They result from the unstable burning of hydrogen and helium on the surface of the neutron star. The sharp increase in X-ray intensity is followed by an exponential decay. The rise occurs in $\sim$0.5--5 s and the decline in $\sim$10--100 s \citep{Lewin1977,Lewin1993,Galloway2008,Bhattacharyya2010,Bhattacharyya2021}.

X-ray bursts have been found to recur on timescales of hours or days. This recurrence need not be at regular intervals. Burst occurring within 45 minutes of the preceding burst is conventionally labelled as short waiting time (SWT) burst \citep{Boirin2007,Keek2010}. Recurrence time as short as 3.3 minutes has been observed in the 11 Hz pulsar, IGR J17480\ensuremath{-}2446 \citep{Motta2011,Linares2011,Linares2012}. Several models have been proposed to explain the occurrence of short recurrence bursts \citep{Fujimoto1987,Taam1993,Boirin2007,Keek2010}. Recently, \cite{Keek2017} proposed that a fraction of the fuel from the previous outburst is transported by convection to the ignition depth, producing a short recurrence time burst. \cite{Beri2019} found that the recurrence timescales observed match that obtained by \cite{Keek2017} but the predicted profile of triple X-ray bursts do not match that observed.

Narrow but high-frequency ($\sim$300-600 Hz) features either during the rising/decay phase or in both phases have been observed in the power spectrum of many thermonuclear X-ray bursts. These features are termed as burst oscillations and arise from brightness asymmetry modulated by the rotation of the neutron star. Hence, oscillation frequency matches the stellar spin frequency \citep[see e.g.,][]{Strohmayer1996,Watts2012,Bhattacharyya2021}. 

\lmxb\ is a neutron star LMXB with a main-sequence companion star, located at a distance of $\sim$6 kpc \citep{Galloway2006}. On the basis of the track traced in the color-color diagram (CCD), \lmxb\ is classified as an atoll source \citep{Schulz1989}. As the source transitions through CCD from a hard state to a soft state, the accretion rate is believed to increase \citep{Bloser2000}. A prolific X-ray burster, \lmxb\ shows all kinds of SWT bursts viz. doublets, triplets and even quadruplets \citep{Ohashi1982,Pedersen1982,Keek2010,Beri2019} during the hard state \citep[see e.g.,][]{Beri2019}. This source exhibits burst oscillations at $\sim$581 Hz, the spin frequency of the neutron star \citep[see e.g.,][]{Zhang1997,Strohmayer1998,Bilous2019,Roy2021}. Although burst oscillations are predominant during the soft state, they also occur during the hard state \cite[see e.g.][]{Galloway2008,Galloway2017}. For \lmxb\ the shortest recurrence time observed is $\sim$5.4 minutes \citep[see][]{Linares2009,Beri2019}.

Here in this paper, we report timing and spectral study of \lmxb\ using data from \astrosat\ and \nus\ observatories. This paper is organised as follows. The observational data and procedure are described in the following section (Section 2). Section~3 is focused on the details of spectral and timing analysis. In Section~4, we discuss the results obtained.
 
\section{Observation and data reduction}
\label{sec-obs}

\subsection{\nus}
\label{sec-nus}

The Nuclear Spectroscopic Telescope Array (\nus; \cite{Harrison2013}) consists of two detectors: Focal plane modules A and B, or FPMA and FPMB), operating in the energy range 3-78 keV. The detectors have a time resolution of 10 $\mu$s and a dead-time of $\sim$2.5 ms \citep{Bachetti2015}. \lmxb\ was observed with the \nus\ on April 27, 2019 (labelled Obs 1 in Table \ref{obs-details}). \cite{Mondal2021} have used the same observation for performing a broadband spectroscopy during the persistent emission of the source. We use the \nus\ data analysis software (\texttt{NuSTARDAS v1.8.0}) distributed with \texttt{HEASOFT v6.26.1} and the latest calibration files (\texttt{CALDB v20201101}) for data reduction and analysis. The calibrated and screened event files are generated using \texttt{nupipeline v0.4.6}. A circular region of radius 100$''$ centered at the source position is used to extract the source light curve and spectrum. Background light curve and spectrum are extracted from the same-sized radial region, away from the source. The task \texttt{nuproducts} is used to generate the spectra and response files.

\subsection{\lxp}
\label{sec-lxp}

Large Area X-ray Proportional Counter (\textsc{LAXPC}) unit onboard \astrosat\ consists of three detectors: \textsc{LAXPC10}, \textsc{LAXPC20}, and \textsc{LAXPC30}, covering the 3-80 keV band with a time resolution of 10 ${\mu}{\rm{s}}$ and a dead-time of approximately 43 ${\mu}{\rm{s}}$. We reinvistigate two of the \astrosat\ observations (labelled Obs 2 and Obs 3 Table \ref{obs-details}) using the \textsc{LAXPC} software (\texttt{laxpcsoftv3.0}) \citep{Antia2017}. These observations have also been reported in \cite{Roy2021} and \cite{Kashyap2021}. Since \textsc{LAXPC30} had ceased operation due to gain instability, it is not used for Obs 2 and Obs 3. Besides, we only use data from \textsc{LAXPC20} for Obs 3 as \textsc{LAXPC10} had very low gain during this observation.


\begin{table*}[htb]
\tabularfont
\begin{center}
\caption{Observation details of \lmxb.}
\renewcommand{\arraystretch}{1.1}
\begin{tabular}{ccccccc}
\topline
Observation ID & Observation & Observation & Exposure & Bursts & Burst labels & Instrument\\
& labels & date & (ks) & & &\\
\hline
30401014002 & Obs 1 & 2019-04-27 & 92 & 15 & B1--B15 & \textsc{FPMA, FPMB}\\
\hline
9000002084 & Obs 2 & 2018-05-09 & 11 & 2 & B16--B17 & \textsc{LXP10, LXP20}\\
9000002316 & Obs 3 & 2018-08-19 & 104 & 14 & B18--B31 & \textsc{LXP20}\\
\hline
\end{tabular}
\label{obs-details}
\end{center}
\end{table*}

\section{Results}
\label{sec-results}

\begin{figure}
\centering
\includegraphics[width=\columnwidth]{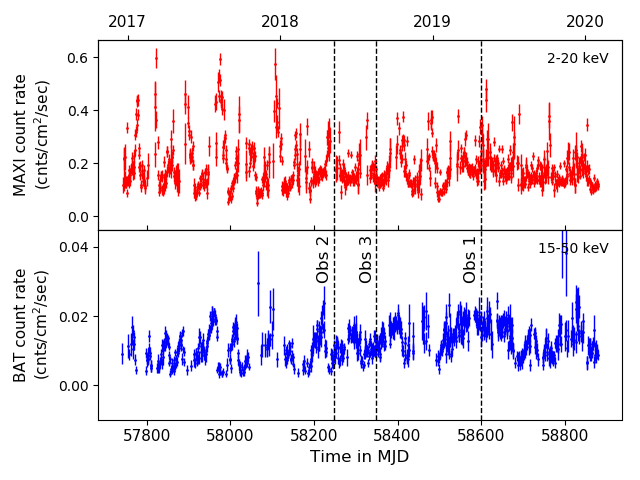}
\caption{\textit{MAXI}-\textsc{GSC} and \textit{Swift}-\textsc{BAT} light curves of \lmxb\ in 2--20 keV band and 15--50 keV band respectively with vertical dash line indicating the observations used in this work. A signal-to-noise ratio of 3 is used as a filter for both the light curves.}
\label{batlc}
\end{figure}

\subsection{\emph{MAXI}-\textsc{GSC} and \emph{Swift}-\textsc{BAT} Light Curves}
\label{sec-batlc}
The hard X-ray transient monitor, \textit{Swift}\,/\,Burst Alert Telescope (\textsc{BAT}), observes $\sim$\,88\% of the sky daily with a detection sensitivity of 5.3 mCrab and a time resolution of 64 seconds in the energy band 15-50 keV \citep{Krimm2013}. The Gas Slit Camera (GSC: \cite{Mihara2011}) onboard the Monitor of All-sky X-ray Image (MAXI: \cite{Matsuoka2009}) covers $\sim$\,85\% of the sky per day with a detection sensitivity of 15 mCrab in the 2–20 keV band \citep{Sugizaki2011}. We use the \emph{MAXI} and \emph{Swift}-\textsc{BAT} light curves of \lmxb\ to infer its spectral state (Figure \ref{batlc}). Both the light curves of \lmxb\ are binned with a binsize of 1~day. A signal-to-noise ratio of 3 is used as a filter. If the source is observed close to the peak of the \emph{MAXI} light curve in the 2-20~keV band, it suggests a soft spectral state of the source. On the other hand, if the source is observed at times close to the peak of the \textsc{BAT} light curve in the 15-50 keV band, a hard spectral state is indicated. From Figure \ref{batlc}, it is clear that Obs 1 and Obs 3 are in hard state. However, the state of Obs 2 is not obvious.

\subsection{Color-Color Diagram}
\label{sec-ccd}

\begin{figure}[!t]
\includegraphics[width=.95\columnwidth]{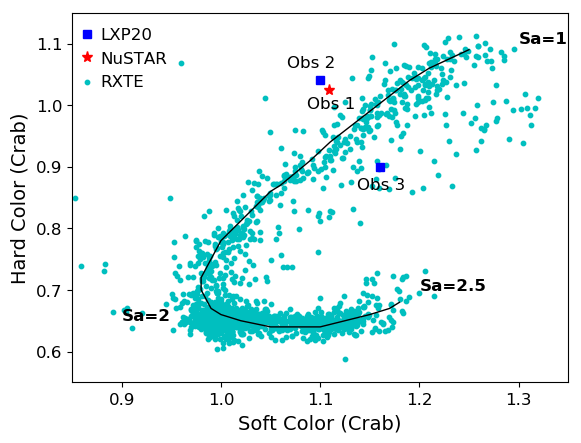}
\caption{Color-color diagram of \lmxb\ with position of the observations studied in this paper against the \rxte\ observations (adapted from \cite{Altamirano2008a}) shown as cyan dots. The thin solid curve, $S_a$, parametrizes the position of the source on the diagram.}
\label{ccd}
\end{figure}

To know the spectral state of the \nus\ and \astrosat\ observations, we plot the color-color diagram (Figure \ref{ccd}) of all the \rxte\ observations of \lmxb\ as given in \citep{Altamirano2008a} and the location of the observations studied in this paper. For the \rxte\ observations, hard and soft colors are taken as the 9.7-16.0/6.0-9.7 keV and 3.5-6.0/2.0-3.5 keV count rate ratios, respectively. The Obs 1 colors are normalized by the corresponding Crab Nebula color values that are closest in time (Obs ID / Obs Date: 10502001010 / 2019-05-09) to correct for the gain changes. For Obs 2 and Obs 3, the nearest \astrosat\ Crab observations (Obs ID / Obs Date: 9000002026 / 2018-04-08 and 9000002364 / 2018-09-13) are used for normalisation. The definition of hard and soft colors for both \nus\ and \textsc{LAXPC} is same as that of \rxte. The counts from both FPMA and FPMB are added to obtain the color values of the \nus\ observation.

The color-color diagram of atoll sources has an island branch and a banana branch which are, respectively, spectrally harder and softer states \citep{Altamirano2008b}. The length of the curve, $S_a$ parametrizes the location of the source on the diagram. $S_a$ is normalized so that it varies from 1 to 2.5 as shown in Figure \ref{ccd} \citep[see, e.g.][]{Mendez1999}, and indicates the mass accretion rate \citep{Hasinger1989}. We find that all the observations belong to the hard spectral state of the source with $S_a\sim1.3$-$1.4$.

\subsection{Energy-resolved burst profile}
\label{sec-enres}

\begin{figure}
\centering
\includegraphics[width=\columnwidth]{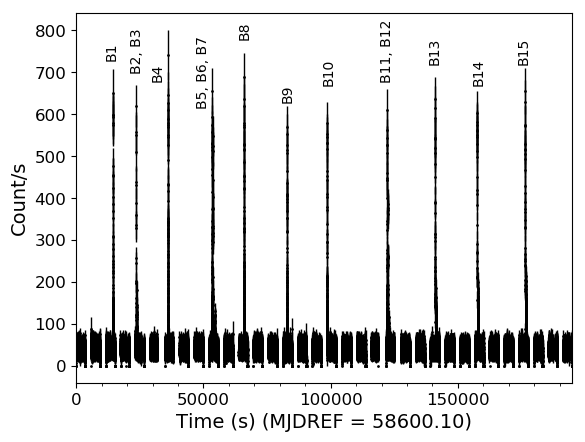}
\includegraphics[width=\columnwidth]{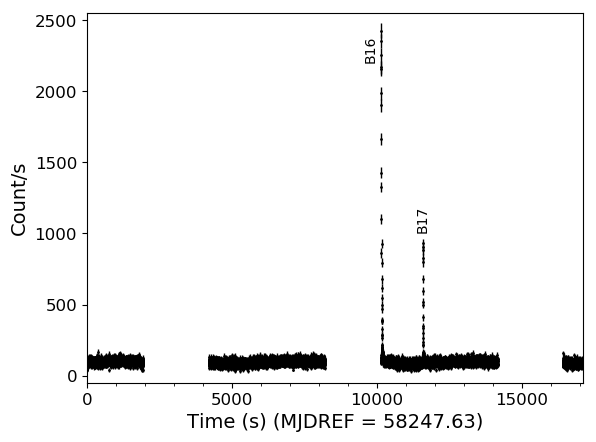}
\includegraphics[width=\columnwidth]{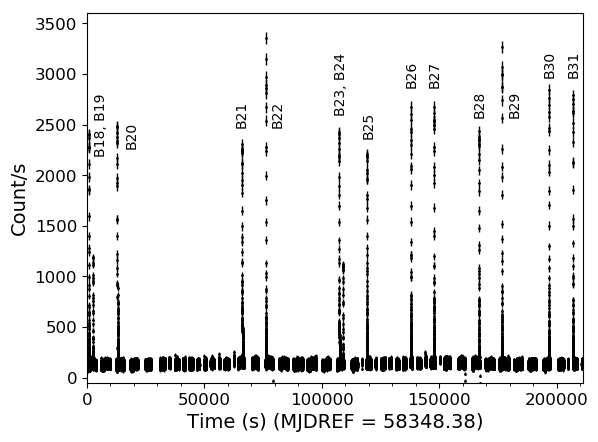}
\caption{\textit{Top panel}: 3-78 keV 1 s-binned \nus-FPMA light curve of Obs 1 containing 15 bursts. \textit{Middle panel}: 3-80 keV 1 s-binned \astrosat-LAXPC20 light curve of Obs 2 containing 2 bursts. \textit{Bottom panel}: 3-80 keV 1 s-binned \astrosat-LAXPC20 light curve of Obs 3 containing 14 bursts. All bursts are indicated by their respective labels.}
\label{full-lc}
\end{figure}

\begin{figure}[!t]
\centering
\includegraphics[width=0.95\columnwidth]{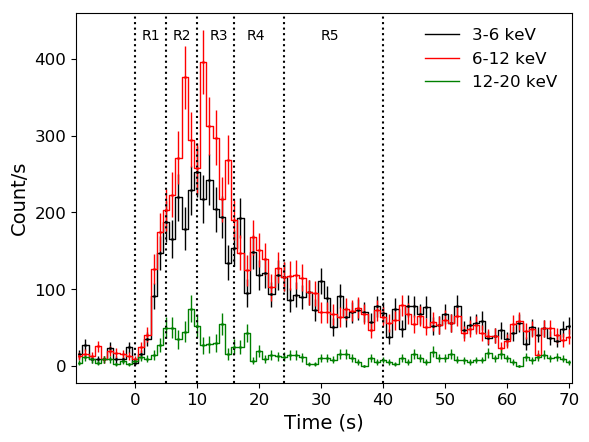}
\caption{3-78 keV 1 s-binned energy resolved \nus-FPMA light curve of burst B1. The dotted vertical lines show the five intervals (labelled as R1-R5) used for time-resolved spectroscopy.}
\label{en-lc}
\end{figure}

We find 15 thermonuclear X-ray bursts (labelled as B1--B15) in the \nus\ observation (see Figure \ref{full-lc}). Their energy-resolved profiles indicate that all bursts are significantly detected up to 20 keV. To illustrate this, we show here the burst profile of one X-ray burst (see Figure \ref{en-lc}). We use three narrow energy bands, viz. 3–6 keV, 6–12 keV, 12–20 keV. In some bursts, such as B1, B7, B10 and B15, a dip is found in the 3-6 keV or 6-12 keV or both bands during the burst peak.

In the two \astrosat\ observations, 16 bursts are found altogether, 14 of which are in Obs 3 alone (see Figure \ref{full-lc}). Bursts B22, B28 and B31 show a dip during the peak. These bursts (B16-B31) are significantly detected up to 25 keV \citep{Roy2021,Kashyap2021}.

We measure the rise time ($t_\text{rise}$) for all the bursts. Rise time is defined to be the time it takes the burst flux to reach 90\% of the peak flux from a time (start time) when the flux was 25\% of the peak value \citep{Galloway2008}. The rise time is found to lie between 2.5 and 8.0 s.

\subsection{Five doublets and one triplet}

\begin{figure*}[!t]
\includegraphics[width=0.66\columnwidth]{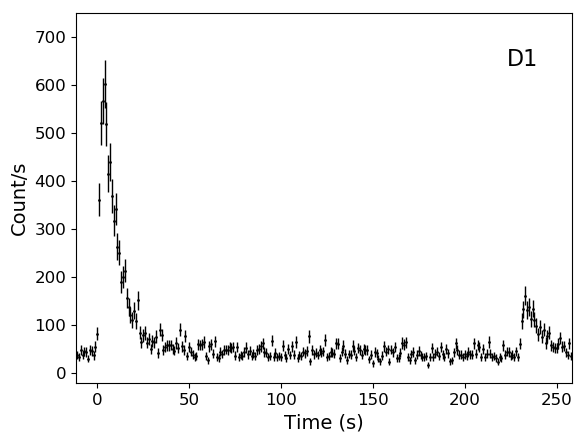}
\includegraphics[width=0.66\columnwidth]{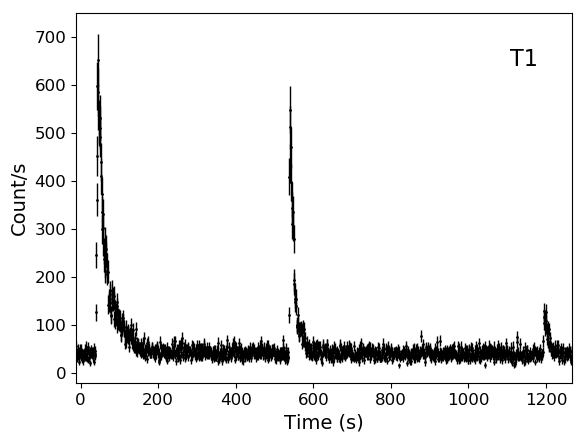}
\includegraphics[width=0.66\columnwidth]{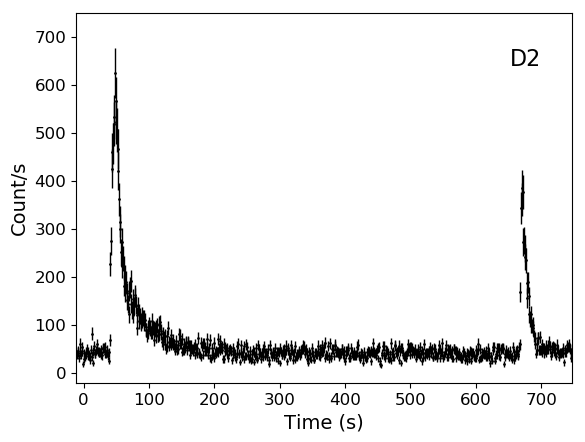}\\
\includegraphics[width=0.66\columnwidth]{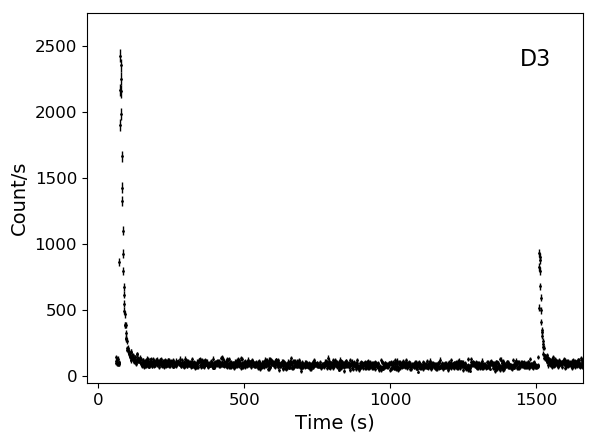}
\includegraphics[width=0.66\columnwidth]{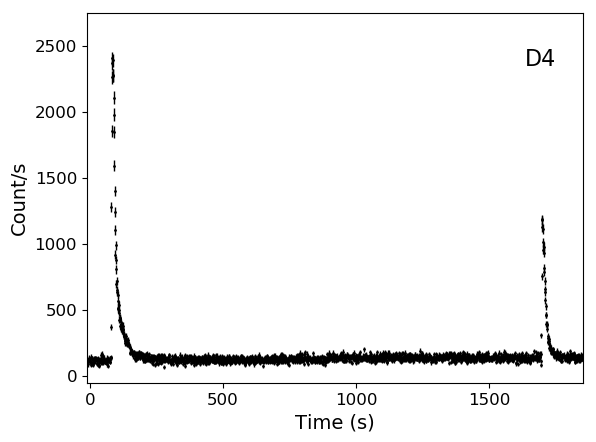}
\includegraphics[width=0.66\columnwidth]{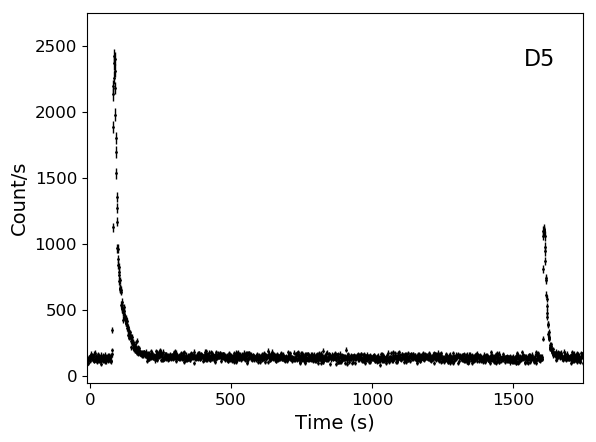}
\caption{\textit{Top row}: 3-78 keV 1 s-binned light curves created using data from \nus-FPMA showing doublet D1, Triplet T1 and Doublet D2 respectively. \textit{Bottom row}: 3-80 keV 1 s-binned \lxpp\ light curves showing doublets D3, D4 and D5 respectively.}
\label{dtd}
\end{figure*}

Among the 15 X-ray bursts, we find multiple short-recurrence bursts (see Figure \ref{dtd}) viz. two doublets, D1 (bursts B2 and B3) and D2 (bursts B11 and B12) and one triplet T1 (bursts B5, B6 and B7). To estimate the recurrence time or wait time between two X-ray bursts, we take wait time to be the separation between their respective peak times \citep{Boirin2007}. The wait time between B2 and B3 in doublet D1 is 3.8 minutes, whereas that between B11 and B12 in doublet D2 is 10.4 minutes. The wait time between B4 and B5 in triplet T1 is 8.2 minutes, and that between B5 and B6 in the same triplet is 10.9 minutes. The rise time for the SWT bursts in these events varies from 2.5 to 3.5 s.

We also find three doublets in Obs 2 and Obs 3. The pair of bursts B16 and B17 in Obs 2 is a doublet (D3), the wait time time being 23.9 minutes. In Obs 3, the doublets are: bursts B18 and B19 (D4) with wait time 26.9 minutes, and bursts B23 and B24 (D5) with wait time 25.4 minutes. The rise time for the short-recurrence bursts in the two observations lies in the range 3--4 s.
 
\subsection{Time-averaged spectroscopy}
\label{sec-tas}

\begin{figure}[!t]
\includegraphics[width=\columnwidth]{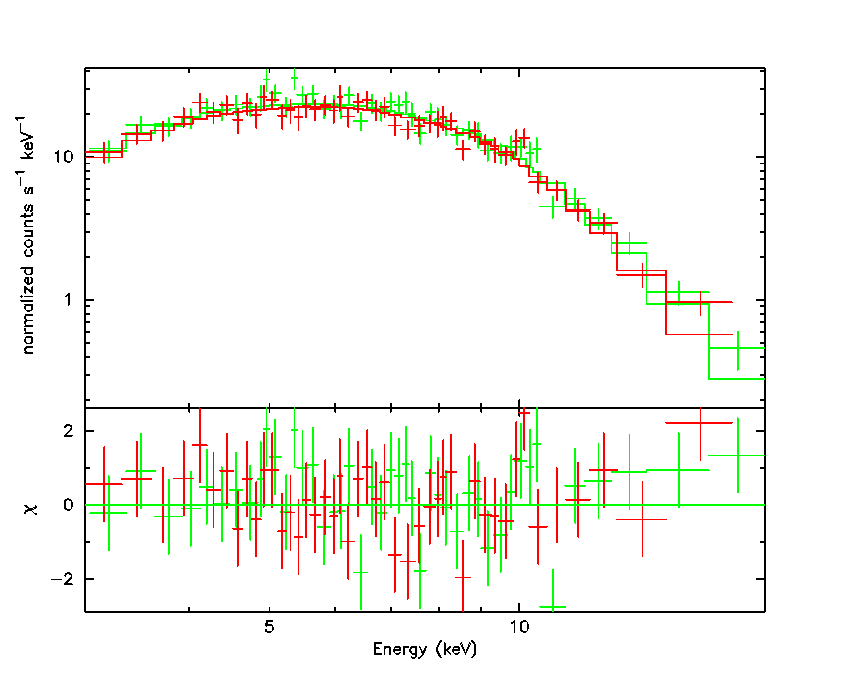}
\caption{\textit{Top panel}: Burst spectrum for burst B1 observed with \emph{NuSTAR}, fitted with \emph{tbabs*bbodyrad} model. The $\chi^2$ (d.o.f) is 0.96 (150). \textit{Bottom panel}: Residues between data and model. The red and green colors correspond to FPMA and FPMB, respectively.}
\label{tas}
\end{figure}

\begin{figure}[!t]
\includegraphics[width=\columnwidth]{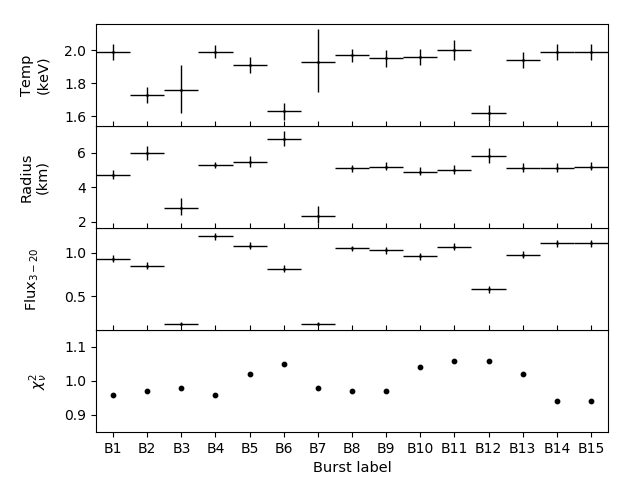}
\caption{Spectral parameters of the bursts in Obs 1 with \emph{tbabs*bbodyrad} model \big(Energy: 3–20 keV; \emph{tbabs}: $n_\text{H}=0.25\times10^{22}$ cm$^{-2}$\big). Unabsorbed flux in 3--20 keV band is given in units of $10^{-8}$ erg~cm$^{-2}$~s$^{-1}$. Burst B3, B7 and B12 are the concluding bursts of doublet D1, triplet T1 and doublet D2, respectively.}
\label{fit}
\end{figure}

For performing time-averaged spectroscopy, we use 100~s preburst emission as the background. The burst duration is taken up to the point where the count rate drops to $1/e$ of its peak value. The 3--20 keV burst spectra are fitted in \texttt{XSPEC version 12.10.1f} \citep{Arnaud1996} with an absorbed blackbody model i.e. \emph{tbabs}*\emph{bbodyrad} (see Figure \ref{tas}). The \emph{tbabs} component accounts for the interstellar extinction. The hydrogen column density parameter, $N_\text{H}$, in \emph{tbabs} is set to $0.25\times10^{22}$ cm$^{-2}$ \citep{Asai2000}. The component \emph{bbodyrad} has the parameters: color temperature, $kT_{\text{bb},\infty}$, and normalization, $K_{\text{bb},\infty}=(R_{\text{bb},\infty}/d_{10})^2$, where $R_{\text{bb},\infty}$ is the apparent blackbody radius in km and $d_{10}$ is the source distance in units of 10 kpc. To account for cross-calibration between FPMA and FPMB, a multiplicative constant component is included in the model. This constant is fixed to 1 for FPMA and allowed to vary for FPMB. The uncertainties correspond to 90\% confidence level. The $\chi_\nu^2$ of the spectral fit ranges from 0.94 to 1.06. The temperature ($kT_{\text{bb},\infty}$) is in the range 1.6-2.0 keV. The radius, $R_{\text{bb},\infty}$ lies between 2.3 and 6.8 km. The unabsorbed flux in the 3--20 keV band, estimated using the \emph{cflux} convolution model, varies from 0.19 to 1.19 $\times 10^{-8}$ erg~cm$^{-2}$~s$^{-1}$. The cross-calibration constant is found to vary in the range 0.95 to 1.11 except for burst B7 for which the constant is found to be 0.77. The best-fit spectral parameters are shown in Figure \ref{fit}.

We try adding an additional blackbody component to the spectra and found a significant improvement in the spectral fits.
We also try replacing the second blackbody component with a powerlaw and in certain cases found a better fit. However, double blackbody model is not well constrained in most cases. Table~\ref{fit-par} gives the best-fit values for selected bursts, and the F-test probability values to indicate the significance of the additional component. The given three bursts of which B4 is the brightest, indicate the presence of an additional component.

The idea that the persistent emission can evolve during an X-ray burst was explored in detail by \cite{Worpel2013,Worpel2015}. They proposed a dimensionless $f_a$ parameter to quantify the same. In this scheme, following \cite{Roy2021}, we fit the preburst emission of 100 s with the model {\emph{`tbabs*(bbodyrad+constant*nthcomp)'}} and then keep the best-fit parameters constant while fitting the burst spectra. The fitting parameters are listed for three bursts in Table~\ref{fit-par}. We find that the spectral fit does improve with $\chi_\nu^2$ ranging from 0.75 to 1.05. The temperatures are in the range 1.5--2.0 keV. The radii range from 1.9 to 6.6 km. The $f_a$ parameter values lie between 1.5 and 4.1. The unabsorbed flux (3--20 keV) varies from 0.33 to 1.28 $\times 10^{-8}$ erg~cm$^{-2}$~s$^{-1}$.

\begin{table}[htb]
\tabularfont
\caption{Spectral fit parameters for time-averaged spectroscopy for selected bursts. ``(f)'' indicates a ``frozen'' parameter. $kT_\text{e}$, $kT_\text{bb}$ and $kT_\text{seed}$ are stated in units of keV.  Refer \S~\ref{sec-tas} for details.}
\renewcommand{\arraystretch}{1.25}
\setlength{\tabcolsep}{8pt}
\begin{tabular}{cccc}
\topline
Parameters & B1 & B4 & B10\\
\hline
& \multicolumn{3}{c}{\texttt{bbodyrad}}\\
\hline
$kT_\text{bb}$ & $1.99_{-0.05}^{+0.05}$ & $1.99_{-0.04}^{+0.04}$ & $1.96_{-0.05}^{+0.05}$\\
Norm & $61.1_{-5.9}^{+6.5}$ & $78.9_{-6.1}^{+6.5}$ & $66.9_{-6.7}^{+7.4}$\\
$\chi_\nu^2$ (d.o.f.) & 0.96 (150) & 0.96 (207) & 1.04 (149)\\
\hline
& \multicolumn{3}{c}{\texttt{bbodyrad+powerlaw}}\\
\hline
$kT_\text{bb}$ & --- & $1.99_{-0.10}^{+0.10}$ & $1.95_{-0.13}^{+0.15}$\\
Norm (bb) & --- & $67.4_{-11.7}^{+17.6}$ & $54.0_{-13.6}^{+22.1}$\\
$\Gamma$ & --- & $2.00_{-0.84}^{1.00}$ & $1.98_{-0.93}^{1.03}$\\
Norm (po) & --- & $0.70_{-0.61}^{+2.06}$ & $0.77_{-0.70}^{+2.54}$\\
$\chi_\nu^2$ (d.o.f.) & --- & 0.90 (205) & 0.93 (147)\\
F-test prob. & --- & 0.0002 & 0.0001\\
\hline
& \multicolumn{3}{c}{\texttt{bbodyrad+fa*nthcomp}}\\
\hline
& \multicolumn{3}{c}{preburst spectrum fitting}\\
\hline
$\Gamma_\text{nth}$ & $1.76_{-0.04}^{+0.04}$ & $1.84_{-0.07}^{+0.07}$ & $1.93_{-0.09}^{+0.11}$\\
$kT_\text{e}$ & 6.89 (f) & 8.59 (f) & 20.0 (f)\\
$kT_\text{seed}$ & 0.30 (f) & 0.20 (f) & $0.75_{-0.26}^{+0.17}$\\
Norm & $0.17_{-0.02}^{+0.02}$ & $0.27_{-0.04}^{+0.04}$ & $0.06_{-0.02}^{+0.06}$\\
$\chi_\nu^2$ (d.o.f.) & 0.93 (114) & 0.82 (115) & 1.14 (122)\\
\hline
& \multicolumn{3}{c}{burst spectrum fitting}\\
\hline
$kT_\text{bb}$ & $1.96_{-0.06}^{+0.06}$ & $1.95_{-0.04}^{+0.04}$ & $1.91_{-0.06}^{+0.06}$\\
Norm & $54.2_{-6.7}^{+7.1}$ & $67.5_{-6.8}^{+7.1}$ & $54.4_{-7.8}^{+8.3}$\\
$f_a$ & $2.20_{-0.67}^{+0.67}$ & $2.56_{-0.65}^{+0.65}$ & $2.76_{-0.68}^{+0.68}$\\
$\chi_\nu^2$ (d.o.f.) & 0.89 (149) & 0.90 (206) & 1.02 (148)\\
\hline
\end{tabular}
\label{fit-par}
\end{table}

\subsection{Time-resolved spectroscopy}
\label{sec-trs}

Since photospheric radius expansion \citep[see e.g.,][]{Tawara1984,Galloway2008} can cause apparent dips in some light curves \citep{Grindlay1980,Bhattacharyya2006a,Bult2019}, we perform time-resolved spectroscopy to check for the same in the bursts observed with \nus. For this, we extract the spectra for the five intervals as shown in Figure~\ref{en-lc} for the bright bursts (B1--B15 except B3 and B7) observed with \nus. The spectra are grouped using \texttt{grppha} to ensure a minimum of 25 counts per bin. We use the spectrum of 100 s preceding the burst to serve as the background spectrum for all the five intervals. The resulting burst spectra are fitted in \texttt{XSPEC} with the model: \emph{tbabs}*\emph{bbodyrad} along with a multiplicative constant component to account for cross-calibration between FPMA and FPMB.

The peak blackbody temperature is found to be in the range 1.7--2.2 keV. The maximum blackbody radius lies between 5.8 and 7.8 km. The unabsorbed bolometric flux (0.1--100~{\rm{keV}}), given by (for each interval): $F_{\text{bol}}=\sigma T^4(R_{\text{bb}}/d_{10})^2=1.0763\times10^{-11}T_{\text{bb}}^4K_{\text{bb}}$~erg\,cm$^2$ s$^{-1}$ \citep{Galloway2006}. The peak $F_{\text{bol}}$ lies in the range 0.8 and 2.1 $\times 10^{-8}$ erg~cm$^{-2}$~s$^{-1}$. We do not infer any photospheric radius expansion in these bursts.

We also use the $f_a$ method (see \S~\ref{sec-tas}) for the bright bursts. The peak temperatures are in the range 1.7--2.2 keV. The maximum radii range from 5.4 to 7.4 km. The peak $F_{\text{bol}}$ varies from 0.8 to 2.0 $\times 10^{-8}$ erg~cm$^{-2}$~s$^{-1}$. The $f_a$ parameter values lie between 0.1 and 8.4.

For the case of \astrosat\ observations (Obs 2 and Obs 3) time-resolved burst spectroscopy has been reported in \cite{Roy2021} and \cite{Kashyap2021}, therefore, we do not perform spectral analysis of the same. For the doublets (D3, D4 and D5) observed with \emph{AstroSat}, none of the bursts show characteristics of radius expansion in the time-resolved spectroscopy.

The characteristic timescale, $\tau$, for the bursts is calculated as $\tau=E_\text{b}/F_\text{pk}$ where $E_\text{b}$ is the burst fluence and $F_\text{pk}$ is the peak flux during the burst \citep[see e.g.,][]{Galloway2008}. For the bursts in Obs 1, this timescale is estimated to be 11--20 s. The $\tau$ values for the bursts in D3, D4 and D5 lie in the range 11--15 s.

\subsection{Power Spectrum: Burst Oscillations}
\label{sec-bo}

\begin{figure*}[!t]
\includegraphics[width=0.95\columnwidth]{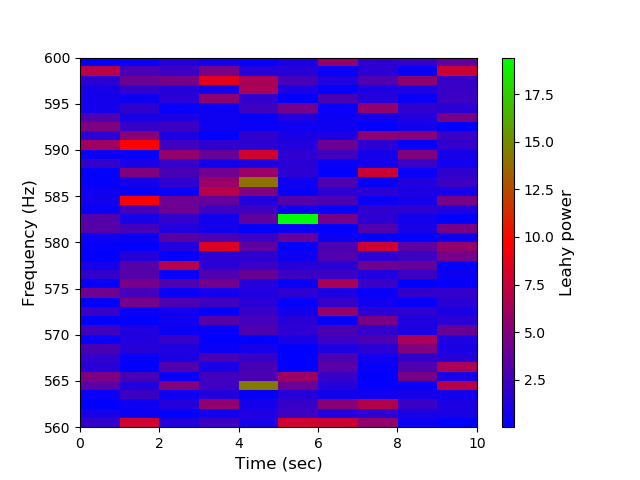}
\includegraphics[width=0.95\columnwidth]{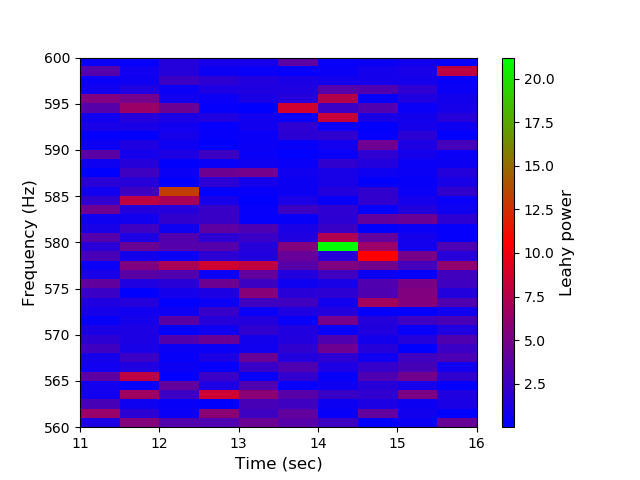}
\caption{\textit{Left panel}: Dynamic power spectra in the 3--20 keV band for 10 s window during burst B2. Each segment is 2 s long and overlaps the previous one by 1 s. (b) \textit{Right panel}: Dynamic power spectra in the 3--20 keV band for 5 s window during burst B10. Each segment is 1 s long and overlaps the previous one by 0.5 s. Burst oscillation is seen at 582 Hz and 579 Hz in B2 and B10 respectively.}
\label{dps}
\end{figure*}

We search for <1024 Hz (Nyquist frequency) oscillations in each of the 15 bursts in Obs 1. Events from both the FPM detectors are combined to reduce the effect of deadtime. The events (photon arrival times) in the 3–20 keV energy band are sampled at the Nyquist rate of 2048 Hz. We perform fast Fourier transform (FFT) of successive 1 s segment (shifting the 1 s time window) of the input barycentre-corrected merged event file corresponding to the burst time interval. The FFT scan is repeated with the start time offset by 0.5 s. A clear signal around 581 Hz is seen during two of the bursts in the Leahy-normalized \citep{Leahy1983} power spectrum. In such bursts, we examine the region that show oscillation and attempt to maximize the measured power, $P_m$, by varying the start and end points of the segment in steps of 0.1 s and trying segment lengths of 1 s and 2 s within a time window of 3 s (20+10=30 overlapping segments). Thus, the number of trials is 30. In both cases, oscillations are found in the decay phase at a confidence level of above $3\sigma$. The results are summarized in Table \ref{bo-details}. The dynamic power spectra are shown in Figure \ref{dps}.

A similar search is performed for the bursts in Obs 3 as well using events from \textsc{LXP20} alone. No clear evidence of burst oscillation is found in them. Previously, \cite{Roy2021} reported burst oscillations in B16 in Obs 2 during the rising phase. We reiterate that result in Table \ref{bo-details}.

To establish the significance of detection of oscillations in burst B2, we carry out Monte Carlo simulations. We generate Poisson-distributed events for the same window size (i.e. 2 s). A deadtime of 2.5 ms is used and implemented by removing any event that occurs within 2.5 ms after a previous event. The initial number of events is chosen such that after the deadtime correction, the number of remaining events is equal to the actual counts within Poisson fluctuations. Two such sequences of events are created to represent the two independent \nus\ detectors, and then merged. We generate 50000 trial windows and calculate corresponding FFTs. The chance probability of occurrence of the observed signal is obtained by counting the number of trial windows with Leahy powers at or above the observed Leahy power. We find three windows with at least one signal above this power in the frequency range 579--582 Hz. Thus, we estimate the chance probability to be $6\times10^{-5}$ which implies a significance of 4.0 $\sigma$ since $X=\sqrt{2}\,\text{erf}^{-1}(1-x)$ where $x$ is the chance probability. For B10, we use a window size of 1 s and obtain a significance of 4.1 $\sigma$ from the chance probability of $4\times10^{-5}$.

Lomb-Scargle periodogram \citep{Lomb1976,Scargle1982, Horne1986}, as implemented in the \texttt{PERIOD} program distributed with the Starlink Software Collection \citep{Currie2014} is also
used to estimate the spin frequency. We obtain consistent values of spin period using this method and the uncertainty estimated corresponds to the minimum error on the period, as given in Table~\ref{bo-details}.

\begin{figure}[!t]
\centering
\includegraphics[width=\columnwidth]{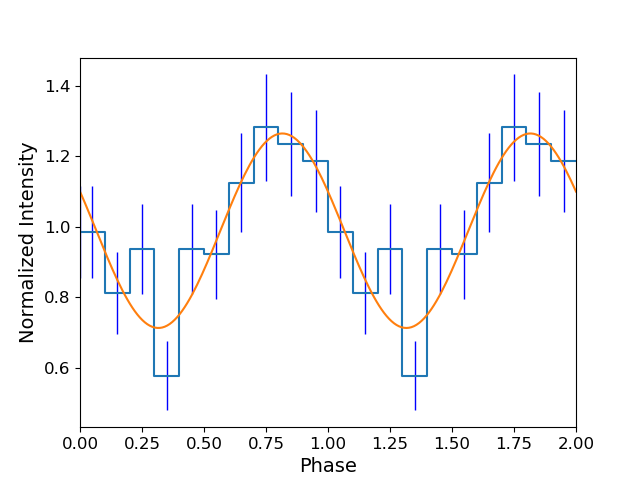}
\caption{Burst profile as a function of phase, plotted in the 3--20 keV band for a 2 s time window during the decay phase of burst B10. The smooth curve shows the sinusoidal fit with frequency 578.81 Hz. The $\chi_{\nu}^2$ (d.o.f.) of the fit is 0.71 (17). The rms fractional amplitude is $19.7\pm2.9$\%. The second cycle is shown for clarity.}
\label{ppm}
\end{figure}

In Figure \ref{ppm}, we show the folded pulse profile for burst B2. The oscillations are modelled with the function: $A+B\sin\,2\pi\nu t$. The rms fractional amplitude is given by $B/(A\sqrt{2})$. The best frequency found with pulse profile modelling \citep{Chakraborty2014,Roy2021} is 582.00 Hz and 578.81 Hz for bursts B2 and B10, respectively. Due to inadequate statistics, we do not study the energy dependence of the burst oscillations found in the \nus\ observation. 

\begin{table}
\tabularfont
\caption{Burst oscillation details. The given frequency values for B2 and B10 are obtained from the \texttt{PERIOD} program. Refer \S~\ref{sec-bo} for details.}
\renewcommand{\arraystretch}{1.1}
\setlength{\tabcolsep}{4pt}
\begin{tabular}{ccccc}
\topline
Burst & Osc. freq. & Phase & Confidence & Fractional\\
& (Hz) & & level & amplitude\\
\hline
B2 & 581.97$\pm$0.13 & decay & $3.1\,\sigma$ & $15.0\pm2.8$\%\\
B10 & 578.75$\pm$0.25 & decay & $3.4\,\sigma$ & $19.7\pm2.9$\%\\
B16$^{[1]}$ & 580 & rise & $4.9\,\sigma$ & $13.8\pm2.2$\%\\
\hline
\end{tabular}
\tablenotes{[1] \cite{Roy2021}}
\label{bo-details}
\end{table}

\section{Discussion and conclusion}

In this work, we present the spectro-timing study of \lmxb\ observed with \nus\ and \astrosat. As seen from the \emph{BAT} and MAXI light curves and from the color-color diagram, the source is in a hard spectral state. This is consistent with that observed by \citet{Mondal2021} (for Obs 1) and \citet{Kashyap2021} (for Obs 2 and Obs 3) who have used the same observation and found that their spectroscopy results favour the hard/low state of the source.

The \nus\ light curve (Obs 1) shows the presence of 15 X-ray bursts including two doublets and one triplet while the \astrosat\ observations~(Obs~2 and Obs~3) show three doublets. So far, six triplets have been reported from this source, including one observed with \emph{AstroSat} \citep{Beri2019}. These authors reported a shortest recurrence time of 5.5 minutes in that triplet, very close to the 5.4 minutes recurrence time reported in a doublet by \cite{Linares2009}. The wait times between two bursts in a triplet seen in this study conform to the wait times in triplets reported in \cite{Beri2019}. The 3.8 minutes recurrence time in a doublet found in our work is the shortest recurrence time reported for this source. To the best of our knowledge, until now only three sources are known to have such a short recurrence time of 3--4 minutes. \cite{Keek2010} reported a recurrence time of 3.8 minutes in 4U 1705\ensuremath{-}44. \cite{Degenaar2012} discovered a recurrence time of 3.8 minutes in SAX~J1747.0\ensuremath{-}2853. IGR~J17480\ensuremath{-}2446 showed a recurrence time of 3.3~minutes \citep{Motta2011,Linares2012}, however, it is considered to be an exceptional source. In the full MINBAR sample of SWT bursts, half of the events arise from a single source, IGR~J17480\ensuremath{-}2446. SWT bursts generally occur at all mass accretion rates, but for some individual sources such as IGR~J17480\ensuremath{-}2446, they may be restricted to a smaller interval of accretion rate \citep[see \S 6.4 of][for details]{Galloway2020}.

\citet{Keek2017} suggest that bursts igniting in a relatively hot neutron star envelope leave a substantial fraction of unburnt fuel at shallow depths. This fuel is brought down to the ignition depth by opacity-driven convective mixing on the observed timescale of minutes and produce a subsequent burst.
Although recurrence timescales observed during these SWT X-ray bursts could be explained by the model proposed by \cite{Keek2017} we do not find any bumps before a follow-up burst in T1 as predicted by these authors. Such differences were also noted by \citet{Beri2019}. \citet{Boirin2007} noted that all sources that show SWT bursts have a fast spinning neutron star ($\nu_\text{spin}>500$ Hz) suggesting the requirement of rapid rotation for multiple-burst events and the importance of rotation-induced mixing processes. They also proposed a waiting point in the chain of nuclear reactions such that a decay reaction with half life similar to the short recurrence times temporarily halts the nuclear~burning. However, \cite{Keek2017} observed that a considerable spread in the recurrence times (3-45 minutes), even for SWT bursts from a particular source, disfavoured this scenario. One plausible explanation to the existence of the short recurrence bursts could be based on the spreading layer model as suggested in \citet{Grebenev2017}. However, these authors also note that further refinement in the model is needed for the complete understanding of the existence of SWT X-ray bursts.

Our time-averaged spectroscopy shows that absorbed blackbody model provides a good fit for the burst spectra yielding an emitting radius ($R_{\text{bb},\infty}$) of $\sim$2.3--6.8 km and blackbody temperature ($T_{\text{bb},\infty}$) of $\sim$1.6--2.0 keV. Similar studies have been carried out by \cite{Degenaar2016} for \lmxb\ SAX J1747.0\ensuremath{-}2853 and their radius and temperature values are in good agreement with ours. However, there is scope for an additional model component such as \emph{powerlaw}, perhaps due to varying temperatures during bursts or plausible deviation from ideal blackbody behavior. We also analyze using the $f_a$ method and find that the spectral fit improves. The $f_a$ values suggest a 
substantial contribution from the persistent emission. Our time-resolved spectroscopy do not show any evidence of photospheric radius expansion \citep[see also][]{Roy2021,Kashyap2021}. However, we would like to add a caveat that the possibility of radius expansion cannot be ruled out due to limited statistics. In most of the bursts, the $f_a$ values are $\gtrsim$5 close to the burst peak suggesting enhanced persistent flux at this stage.

The actual emitting radius ($R_{\text{bb}}$) and blackbody temperature ($kT_{\text{bb}}$) are given by $R_{\text{bb}}=R_{\text{bb},\infty}\cdot f^2/(1+z)$ and $T_{\text{bb}}=T_{\text{bb},\infty}\cdot (1+z)/f$, respectively \citep{Sztajno1985}, where $(1+z)=(1-2GM/Rc^2)^{-1/2}$ is the surface gravitational redshift, $f$ is the color factor which accounts for the scattering of burst photons by the electrons and the frequency-dependent opacity of the neutron star atmosphere, and $R/M$ is the neutron star radius-to-mass ratio. We do not apply the correction factors for the measurement of a real value of neutron star radius.

Bursts which are helium-rich, are expected to have shorter characteristic timescales ($<$10), fast rise times ($\lesssim$2) and peak flux exceeding the Eddington limit \citep[see e.g.,][]{Galloway2008}. The characteristic timescales of 11--20 s, long rise times of 2.5--8 s and the lack of any radius expansion signature indicate that the bursts are hydrogen-rich bursts. This is consistent with the observation that the sources exhibiting short-recurrence bursts are all hydrogen-accretors. In our best knowledge, till date, no SWT burst has shown radius expansion \citep[see e.g.,][]{Galloway2008,Keek2010,Keek2017}.

It is generally believed that the highest burst rates on average are found for the sources with burst oscillations. However, it may be a selection effect as not all bursts exhibit burst oscillations, and therefore, a high burst rate means observation of many bursts and hence burst oscillation detections. 
Oscillations seen in this study are consistent with the range 578-582 Hz given in \cite{Bilous2019} although most of the burst oscillations from this source are seen at 580-581 Hz. We detect a strong signal at a frequency (582 Hz) similar to that observed during one of the superbursts in \lmxb, albeit with a much higher fractional amplitude (see Table-\ref{bo-details}) compared to $\sim$0.01 observed in the superburst. In the case of accretion-powered pulsations in intermittent AMXPs, the pulsation frequency is $\sim$0.5--1 Hz above the burst oscillation asymptotic frequency with fractional amplitude of up to a few percent \citep[see recent review by][]{Bhattacharyya2021}. Therefore, 582 Hz pulsations observed during the 2001 superburst are believed to be accretion-powered \citep[see][for details]{Strohmayer2002}. Given a higher value of fractional amplitude of the oscillations detected in our study and considering an uncertainty up to 0.4 Hz \citep{Muno2002}, the higher value of burst oscillation frequency cannot be ruled out.

During the decay phase of two bursts in Obs 1, the oscillations are found with $\sim3\sigma$ confidence. Decay phase oscillations are generally described through the cooling wakes \citep{Cumming2000} or the surface oscillation modes \citep{Heyl2004}. Cooling wake is the temperature asymmetry due to cooling of the neutron star's surface, spreading across the star in a finite time after an X-ray burst. High amplitude decay oscillations (>10\%) are usually explained using this model \citep{Mahmoodifar2016}. Surface mode is the asymmetry in the neutron star ocean or atmosphere excited during an X-ray burst. The oscillation amplitudes are typically $\sim$10\%. It may be plausible that both the models contribute to the evolution of burst oscillations \citep{Roy2021}.

\section{Summary}

\begin{itemize}

\item~In 207 ks of observation of \lmxb\ made during its hard spectral state with \nus\ and \astrosat, we find five doublets and one triplet. The recurrence time of 3.8 minutes in one of the doublets is the shortest recurrence time reported for this source.

\item Our time-averaged spectroscopy shows that an absorbed blackbody model provides a good fit. The temperatures and radii measured during each of the bursts in Obs 1 are consistent with measurements reported using similar approach.

\item From time-resolved spectroscopy, no radius expansion could be inferred in any of the burst. The characteristic timescales and the rise times suggest that the composition of the bursts is primarily hydrogen.

\item Burst oscillations are found in the decay phase of two of the bursts observed with the \nus. One of these bursts is part of a doublet. The oscillations are found near the known spin frequency (581 Hz) of the source. A 582 Hz burst oscillation is seen in one of the bursts.

\end{itemize}


\normalfont

\section*{Acknowledgements}

AB and PR acknowledge the financial support of ISRO under \emph{AstroSat} archival Data utilization program (No.DS-2B-13013(2)/4/2019-Sec.2). AB is grateful to the Royal Society, U.K. and to SERB (Science and Engineering Research Board), India and is supported by an INSPIRE Faculty grant (DST/INSPIRE/04/2018/001265) by the Department of Science and Technology (DST), Govt. of India. This research has made use of the \nus\ data analysis software (\texttt{NuSTARDAS}) jointly developed by the ASI science center (ASDC, Italy) and the California Institute of Technology (Caltech, USA). The data for this research have been provided by the High Energy Astrophysics Science Archive Research Centre (HEASARC) and the Indian Space Science Data Centre (ISSDC). We thank the anonymous referee for important inputs towards the improvement of the paper.





\appendix



\end{document}